**Automated Design Appraisal: Estimating Real Estate Price Growth and Value at Risk due to Local Development**

Adam R. Swietek[a,*]

[a] *Laboratory of Environmental and Urban Economics (LEURE), École Polytechnique Fédérale de Lausanne (EPFL), Lausanne, Switzerland*
\* *Corresponding author: adam.swietek@epfl.ch*

December 2023

## Abstract

Financial criteria in architectural design evaluation are limited to cost performance. Here, I introduce a method – Automated Design Appraisal (ADA) – to predict the market price of a generated building design concept within a local urban context. Integrating ADA with 3D building performance simulations enables financial impact assessment that exceeds the spatial resolution of previous work. Within an integrated impact assessment, ADA measures the direct and localized effect of urban development. To demonstrate its practical utility, I study local devaluation risk due to nearby development associated with changes to visual landscape quality. The results shed light on the relationship between amenities and property value, identifying clusters of properties physically exposed or financially sensitive to local land-use change. Beyond its application as a financial sensitivity tool, ADA serves as a blueprint for architectural design optimization procedures, in which economic performance is evaluated based on learned preferences derived from financial market data.

# 1    Introduction

In architectural design optimization, computer generated designs are iteratively evaluated with respect to building performance criteria. While building design concepts are commonly assessed for engineering performance, including structural resilience[1], environmental quality[2], [3] and energy performance[4], [5], as well as for cost performance, including material usage[6] and sustainability goals[7], [8], current approaches have stopped short of considering the financial value of a given design directly, or more broadly speaking the preference thereof. The limited feedback between building design and real estate valuation models can be attributed to a lack of availability of simulations and pricing models with similarly specified attributes and parameters, in part due to the traditional separation of disciplines.

Most current studies in real estate economics utilize valuation models in specific geographic regions to infer the marginal price effect, or the price premium, of a given performance metric, including environmental amenities such as streetscape[9], waterscape[10], viewshed[11], building morphology[12], greenery[13], daylighting[14], visual quality[15], and landcover[16]. Yet, to our knowledge, no study has utilized these fitted models to predict the price of newly generated building and urban designs. This would require incorporating pricing models within a generative design or optimization framework or within a risk framework to assess the impact of attribute persistence: An example for the latter would be whether a desirable lake-view is exposed to future obstructions. The challenge arises from the need to additionally generate new building designs, and, for risk assessment studies, to compute the exposure to urban development or land-use change. While parametric design and building simulation are central to architectural design optimization[17], limited access to relevant transaction data and model parameters, hampers efforts to evaluate the preference or value of a generated design, as well its impact on the local context.

To overcome these challenges, this paper introduces an integrated workflow, an augmented valuation model called Automated Design Appraisal (ADA). The ADA algorithm incorporates computational design techniques to generate a city model based on design parameters; geometric computing to simulate building performance; and finally, a fitted econometric model that predicts the value of a building's design. The output is a single value representing the weighted economic preference of the individual attributes defining a single building design concept and its surrounding context. As a structured approach, ADA can be incorporated within various design analytic frameworks, including design optimization, or risk and impact assessment, by perturbing the initial design parameters and subsequently quantifying the effect size of an altered design scenario.

To demonstrate its usefulness, this paper implements ADA within the context of a visual impact and risk assessment. It presents results from two case-studies in Lausanne, Switzerland; (1) the impact due to a single proposed development and (2) the potential value at risk due to nearby land-use changes across an entire commune. Importantly, the effect size is assessed not only at the point of alteration, but also for nearby buildings to capture the imposed cost of a generated design, or put another way, the risk of neighbor property devaluation. The results illustrate the theoretical space of localized costs imposed on the neighborhood due to simulated design scenarios.

To assess both direct and localized effect sizes, it is important to choose an appropriate building performance metric by which to benchmark one urban design against another. Of the environmental performance metrics, a building's view or visual landscape is particularly relevant as high-quality views are considered inherently important to home prices[18], [19]. Moreover, visual obstructions and the subsequent risk of devaluation are primary drivers of objection to proposed developments by community members – a sentiment typically referred to as NIMBYism (**n**ot **i**n **m**y **b**ackyard)[20]. We therefore use a building's Visual Capital (VC), a value that evaluates building level visual landscape quality, in our case study. VC is as an income derived, non-linear weighting of the visible share of landscape elements[21], and, importantly, is derived directly from 3D building geometries and is thus sensitive to nearby design changes to the urban environment. Additionally, we fit a pricing model trained on transaction data, provided by Wüest Partner, learning the preferences for VC and other covariates, and subsequently apply this model to gauge the magnitude of change in predicted price with respect to changes in VC across our design scenarios.

The proposed methodology and impact analysis can be further extended to examine the cost/benefit of proposed urban infrastructure, optimized greenery layouts, as well as its effect on other location-based attributes. The proposed design appraisal offers insights into the direct gain and social acceptance of design choices, making it a tool for site-selection and feasibility assessments. Additionaly, future design optimization studies can leverage ADA, by converting performance metrics into financial metrics, to aggregate building objectives into a single value, producing a preference ranking of the 'optimized' set of designs.

## 2 Literature Review

### 2.1 Economic Performance Metrics

In the context of architectural design, economic performance has been described as the evaluation of revenue, cost, and profitability [17]. Commonly used economic performance metrics focus on a cost minimization objective, such as the cost of pedestrian walking routes [8], or the cost of lighting & heating (or space efficiency) [6], [22]. For example, Nagy et al utilize a profit metric to explore modular design solutions at the urban scale [23]. Using pre-defined values for selling price and project cost for each modular unit type, a generative design procedure produces a set of profit-optimized solutions. However, the approach has specific limitations: When a fixed selling price is applied, it overlooks the significance of the unique spatial qualities within the proposed design. This can contradict the proven value of the design itself [12]. In addition, the potential cost imposed on neighbors as a result of new development [24] remains unexplored.

The few studies that have focused on evaluating the preference of a generated design primarily leverage satisfaction questionnaires [25]–[27]. Such stated preference approaches only describes the hypothetical preference which itself may be biased[28]. In contrast, revealed preference methods, such as the determining the willingness to pay by regressing attributes on transaction prices, describe actual economic decisions [29], are considered a superior method to measure preference.

Thus, the current study contributes to the literature in two ways: it provides a new method that leverages revealed preferences using real estate transaction data to ascribe economic value of newly proposed designs, and it simultaneously estimates the economic impact of a design solution on its immediate urban surrounding. It thus allows to assess the devaluation risk due to land-use change.

### 2.2 Devaluation Risk

Devaluation risk, or potential decrease in the value of a property, is a major concern to property owners and lenders. Previous work primarily focused on the devaluation due to climate change [30]. Typically, the effects of physical risks are estimated by using historical financial and environmental data; where natural disaster shocks, such as flooding [31]–[35] and wildfires [36], are used to show persistent negative impacts on housing values. To understand the future and potential impact of climate change on real estate, the generation of hazard exposure maps is essential. For example, high resolution flood hazard maps for the year 2020-2050 [37] enabled subsequent studies to assess whether residential properties are over-priced relative to their flood exposure [38].

Among the risks to real estate owners is property devaluation due to local land-use change [20]. For example, Thibodeau shows that the development of a high-rise building had a negative effect on the property values of adjacent neighbors (< 2,500 meters) [24]. At such a local scale, it is possible to compute exposure maps by leveraging computational design, urban analytics, and micro-climate simulation methods, including energy modeling [4], solar irradiation [39], daylighting [3], and visibility [40]. Past studies have leveraged these simulations and applied the hedonic pricing model [29] to assess the marginal price effect of micro-climate performance on real estate valuation [15], [41]–[43]. Yet, unlike the future flood risk projections example, local risk evaluation methods stop short of examining the sensitivity of a set of building valuations across future urban design scenarios. Thus, this paper extends the literature by taking advantage of a key feature of geometric data, that differentiates it over other urban data types: it's mutability. Specifically, a sensitivity analysis which can be applied to generate new design scenarios and to automatically assess the impact of design perturbation on property values.

## 2.3 Visibility Simulation and Visual Capital

Of the factors that drive property devaluation risk, visual impact resulting from land-use change is of particular concern to NIMBYs [20], [24]. This concern is driven by the significant influence of attractive views on property values [11], [18], [43], [44] and the localized effect of visual obstructions [24]. Views encapsulate an abstract summary of the urban environment from a single perspective, making it easier for individuals to notice changes in the landscape aesthetics compared to aspects such as noise or air pollution. Yet, despite the importance and attention paid to visual impact assessment[45] and visual landscape research more broadly[43], access to a structured 3D approach to evaluate visual landscape at the building-level has only been achieved recently, in the form of the Visual Capital (VC) index [21].

The computation of the VC index is composed of three essential parts: (1), the viewpoint visual share simulation, (2) a set of aggregation functions defining building view-metrics, and (3) a machine learning model that predicts net-income, with the latter serving as a proxy for economic preference based on the concept of amenity-based income sorting [46], [47]. The viewpoint visual share simulation leverages the raycasting algorithm originating from a set of façade points to determine the 'visible' part of a 3D city model, and recording the attributes of the intersected ray, including distance, obstructions, and landcover category. The generated viewpoint visual share dataset indicates what landcover categories are visible and in what proportion from a single viewpoint, before being aggregated to the building-level. Specifically, viewpoints are grouped by their associated building and a series of aggregation functions are mapped, resulting in a set of 57 view-metrics describing the spatial composition and configuration of visible landcover elements for each building. View-metrics include average sky exposure, maximum visual share of nature, visual access to lake-view, balance of elements in distance, richness of panorama, among others. A neural network then estimates that weighted importance of these building view-metrics in predicting the commune average net-income. And finally, applying the fitted model to out of sample visual share data produces a building's VC index.

Unlike other view-based building performance metrics, VC is a single value and can be easily integrated within pricing models to determine the price-amenity gradient. In addition, it can easily be derived for newly generated design scenarios, thus providing a direct link between design performance evaluation (the view) and pricing.

# 3 Material and Methods

Automated Design Appraisal (ADA) is the application of a fitted pricing model to evaluate the economic preference of multiple design metrics. We demonstrate the applicability of ADA by integrating it within a devaluation risk assessment focused on the potential visual impact of a simulated urban development. The workflow includes three parts: (1) pricing model (2) design simulation and (3) parametric design generator. To identify the potential financial impact, we measure the difference in predicted price between the simulated urban design scenarios (*alt*) and the as-built design scenario (*ref*).

## 3.1 Pricing Model

To analyze the relationship between the design attributes of a generated design scenario and real estate sale transactions, I use the hedonic pricing model. The commonly used approach in real estate economics literature quantifies the revealed preference, or the buyer's willingness to pay, for a given characteristic. These building characteristics includes immutable attributes, including year of transaction, year of construction, etc., as well as mutable attributes, which are the variables of interest within parametric design and design evaluation. Eq. 1 presents the functional form of the specified model,

$$\ln(P)_i = \beta_0 + \beta_1(VC)_i + \beta_3(L)_i + \beta_4(M)_i + \beta_5(S)_i + \beta_6(T)_i + \varepsilon_i \quad (1)$$

where the dependent variable ln(P) is the natural logarithm of the transacted sales prices for building observation *i*. In this paper we are interested in quantifying the price sensitivity with respect to visual impact, thus we use Visual Capital (VC) as the variable of interest. L is a vector of exogenous location characteristic, including the log-scaled distance to water bodies. M is a vector of neighborhood level characteristics, such as macro-location[48]. T is a vector representing time fixed effects, i.e. year of transaction, and $\varepsilon_i$ is a vector of the unobservable characteristics.

    Given the importance of water-bodies on property valuation and on VC, we further limit the training sample to transactions of buildings located within agglomerations in proximity to a major lake, i.e. Biel/Bienne, Zurich, Lausanne, Geneva, Vevey–Montreux, Luzern, Thun, Neuchatel, and Zug. To control for differences between these urban regions of Switzerland, we condition a building's VC on agglomeration identity. Transaction data, including 7,651 sales transactions from years 2008 to 2017, and exogenous data points were provided and anonymized by Wüest Partner in compliance with Swiss privacy laws.

## 3.2 Design Simulation

A building's design performance is measured with respect to its visual landscape quality.

### 3.2.1 City Model

To evaluate a building's visual landscape, I first construct a Digital Twin, or 3D city model, using three separate publicly available databases: representing terrain, buildings, and vegetation [49]–[51]. The composed city model provides a 3D digital representation of the building stock and is used as the reference scenario (*ref*). Importantly, due to the mutability of 3D data we can subsequently alter the input geometries to represent design changes. The swissBUILDING3D database provides separate 3D geometries for a building's facade and roof, which allowing the modification of the height of an individual building, e.g. add a story, without distorting the roof. The altered design parameters thus lead to a slightly modified city model (*alt*). To represent different design scenarios, I compile a set of structured design alterations that can be compared against one another and against the reference scenario.

### 3.2.2 Performance Metrics

Using the compiled city model, a viewpoint visual share dataset is generated and subsequently used to compute a range of view-metrics and the Visual Capital index, as described in Swietek et al [21]. Specifically, I compute viewpoints for $J$ buildings ($B$) indexed by $j$. The $j$-th building has $n$ viewpoints ($B_{jn}$) situated on its façade, spaced apart by 8 meters across each floor. Importantly, only exterior walls are considered. For instance, in the case of two buildings joined by an interior wall (e.g. row of townhomes) they are considered as single joint structure. For each viewpoint $B_{jn}\ with\ n = 1,2,...,n$, a 120-degree view cone composed of 2600 rays is cast outward and the endpoint of intersection within the city model is recorded. The count of rays intersecting the same $l = 1,2,...,20$ landcover categories at $d = 1,2,...,4$ distance categories are summed and divided by the total number of rays(i.e. 2600), generating the visual proportions for $B_{jn}$ denoted by $z_{ld}$. Visible proportions of landcover data for building $j$ are thus represented by a $(nx20x4)$ array, denoted by $Z_j$. The values are derived from the swissTLM3D, COPERNICUS databases, describing whether the view is obstructed by a façade, roof, or vegetation; as well as the distance to visual elements. This procedure is referred to as the viewpoint visual share simulation, or visibility analysis. Next, the generated viewpoint visual share dataset is used to aggregate the land-use proportion viewpoint values to building level view metrics (for details see [21]). This results in 57 view metrics describing the visual landscape for a given building, e.g. maximum share of lake-view, sky exposure, etc. Lastly, to generate the Visual Capital index, I apply the pre-trained neural network, from Swietek et al, to the newly constructed vector of view-metrics.

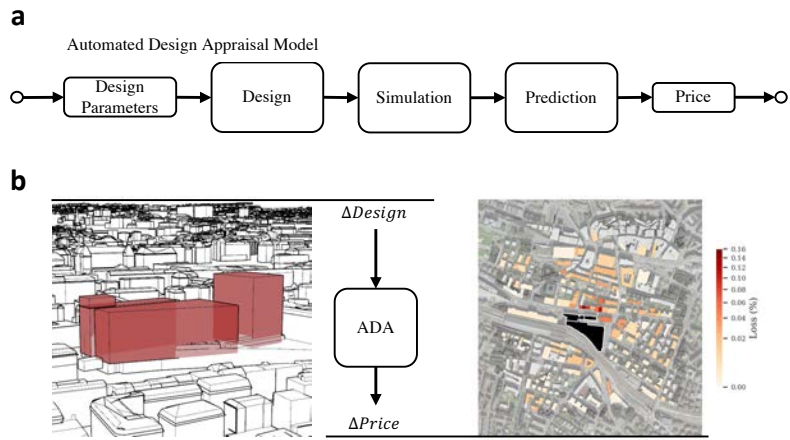

*Figure 1 (a) Abstract schematic of the proposed Automated Design Appraisal algorithm: step 1.) define the set of design parameters of interest; step 2.) update buildings within 3D city model according to design parameters, thereby creating an alternative design scenario; step 3.) compute building performance metrics using building and micro-climate simulations – in this paper, we utilize a viewpoint visual share visibility simulation to generate a set of view-metrics and subsequently calculate the visual landscape quality, i.e. Visual Capital; step 4.) update vector of building attributes to include new performance metrics; step 5.) use fitted model to predict price of building with updated performance metrics. (b) Abstract schematic showing the application of ADA for a visual impact assessment in Lausanne. The proposed development, shown in red, represents the point of modification within a reference 3D city model. The spatial distribution of price impact, computed via ADA, is shown, with darker red representing greater impact on a building's predicted price.*

## 3.3 Integrated Impact Assessment

The determinants of risk are the degree of exposure and sensitivity to a given hazard[52]. In the context of this paper, a hazard is a proposed building development that may obstruct the view and degrade the visual landscape quality of nearby buildings. As such, I propose two case-studies: an impact assessment of a single hazard, and of multiple hazards.

To measure the exposure of building j to changes in the urban form, we can iteratively perturb the underlying city model denoted by $s^{ref}$ thereby creating a set of new design scenarios of length $S$, and measure the persistence of the performance values. The design evaluation procedure consisting of $M$ metrics (here $m = 1, 2, \ldots, 57$ view metrics) applied to building $j$ derived from the context of design scenario $s$, results in a $MxJxS$ matrix $V^{alt}$ of design performance values. To express the impact or change in performance metrics in a given building,

$$\Delta V = V^{alt} - V^{ref} \quad (2)$$

Where $\Delta V(m)$ is a $JxS$ matrix and $V^{ref}(m)$ is a vector of length j, describing the design performance values of metric $m$ in the as-built design scenario $s^{ref}$ across all included buildings $j$. To standardize the impact to represent the relative change,

$$\Delta V^{rc} = \frac{\Delta V}{V^{ref}} - 1 \quad (3)$$

Thus, $\Delta V(m)_{js}^{rc}$ describes the relative change in the value of metric $m$ for building $j$ due to the proposed project s. To express the maximally exposed metric, I take the metric with the largest change for each building and scenario

$$MEVM_j = Max_M \left( \Delta V(m)_{js} \right) \quad (4)$$

To derive the impact on predicted price $\Delta Y$, I take the difference in the predicted prices of building $j$. Where $Y$ is a $JxS$ matrix. The predicted price is calculated by applying the previously development pricing model to the sample region with updated values for building performance values, $V^{alt}$. Further, to identify the financial impact due to the effect on a building's visual capital, I simplify the price impact equation by assuming no change across the other building's attributes. Thus,

$$\Delta Y = Y^{alt} - Y^{ref} \approx \beta_{vc} \Delta V(VC) + \varepsilon \quad (5)$$

### 3.3.1 Single Development

The first case study examines the potential visual impact of the Rasude Development within a .5km radius of a proposed 15-story office project near the Lausanne train station [53], [54]. Thus, one new design scenario $s^{alt}$ is a modified city model containing the proposed Rasude Development. The proposed massing, containing three distinct structures[55], is designed in Rhinoceros 3D and added to $s^{ref}$ replacing the existing structures. Next, the design performance simulation with respect to a building's visual landscape quality is initiated (described in section 3.1.2) and spatial view metrics are calculated for both design scenarios.

## 3.3.2 Regional Vulnerability

The second case study pertains to assessing the risk of multiple hazards, the spatial distribution of vulnerability to land-use changes within a sample region. Unlike the first case study, it incorporates multiple design scenarios and contrasts the potential gain in value of the up-zoned building to the potential losses in value of its neighboring buildings. Specifically, using the process iteratively modifies each building in a sample by adding 1 floor (i.e. 5 meters) to the existing building structure. Thus, in a sample region of 204 buildings index by j, this design augmentation results in 204 alternate design scenarios indexed by $s$. Using this set of design scenarios, we next compute the visibility performance of buildings in the sample region. Importantly, for each iteration, we dynamically limit the sample region to the point of modification and its nearest 9 buildings. This helps to reduce the compute time, while maintaining the buildings expected to be most vulnerable to the change within a sample. As a result of this procedure, 2244 design performance simulation were executed: where in addition to the reference design scenario (no modifications), the 204 buildings were modified and the visual impact of each modification was assessed either from the perspective of the modified building itself or from the perspective of each of the nearest 9 neighboring buildings. This results in a sparse $JxS$ matrix $\Delta V$, where each design scenario corresponds to a specific modified building. Hence diagonal entries of the matrix of $\Delta V$ represent the impact of the modification on the building itself, or direct effect ($DE$). DE is a vector of size J that represent the increase (benefit) in a given metric at the modified site.

$$DE(m) = \Delta V(m), where\ s = j \qquad (6)$$

Whereas the off-diagonal entries represent the impact of a modification on nearby neighbors, defined as local effects ($LE$).

$$LE(m) = \Delta V(m), where\ s \neq j \qquad (7)$$

As LE maintains a two-dimensional representation, we additionally compute a vector of cumulative local effects and exposure to local effects. Cumulative local effects (CLE) illustrative to collective impact of a single modification on its neighboring buildings.

$$CLE_s(m) = \sum_j \Delta V(m)_s, where\ s \neq j \qquad (8)$$

On the other hand, exposure to local effects (ELE; from the perspective of an unaltered neighbor) denotes the maximum change experienced across all design scenarios s. Put another way, this indicates the potential value at risk attributed to simulated land-use changes in the vicinity of a building.

$$ELE(m)_j = Max_s(\Delta V(m)_j), where\ s \neq j \qquad (9)$$

Figure 2 illustrates the spatial distribution via impact maps portraying maxVSH: Sky, the maximum proportion of sky visible from a single viewpoint. Further, an abstract graph network represents the relationship considered across the impact assessment metrics: ref, DE, CLE, and ELE metrics.

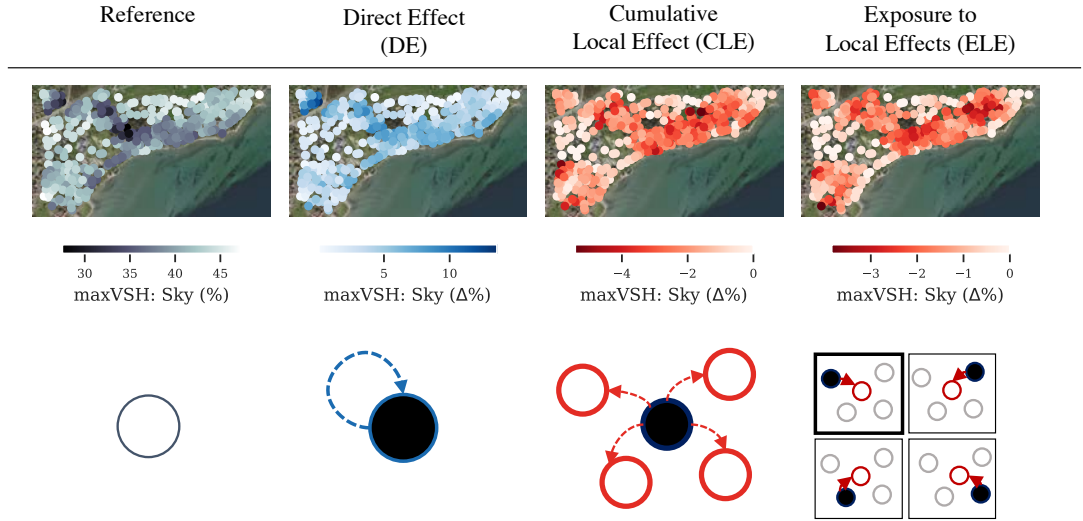

*Figure 2 Spatial distribution and abstract representation of the integrated impact assessment metrics used to visualize the distribution of impacts on maxVSH Sky, i.e. the maximum visible proportion of sky from a single viewpoint across all of a building's viewpoints. Reference is the as-built condition of the city, Direct Effect (DE) express the gain in Sky Exposure as a result of the up zoning, Cumulative Local Effects (CLE) describes the gross cost imposed on its neighbors due up zoning at a given building, and Exposure to Local Effects (ELE) expresses the maximum potential loss across all of the unzoning scenarios tested.*

# 4 Results

## 4.1 Value of a View

We carry out a hedonic regression to understand the net effect of each included variable (see Methods) in predicting the sales transactions of included buildings. We are particularly interested in the coefficients for Visual Capital, as this learned parameter will drive variability across our integrated impact assessment tool. Table 1 shows parameter estimates across four pricing models where the natural logarithm of transacted prices is used as the dependent variable. We test four specified models (Table 1) to understand the interaction of Visual Capital across two different location-based scenarios, fitting VC independent (model 1 and 3) or dependent (model 2 and 4) on the agglomeration buildings are located in. In addition, each location-based model excludes (model 1 and 2) or includes (model 3 and 4) agglomeration and a macro-location indices[48] provided by Wüest Partner. The difference between the first and second model (as well as between third and fourth model) helps identify the variable importance VC has across agglomerations. The third and fourth model additionally control for a set of important covariates, including a macro-location index which describes desirability across communes. Thus, the difference between the third and fourth model, highlights the spatial variability of VC after controlling for both building- and macro-level covariates. Importantly, the ranked coefficients for agglomeration-specific VC remain consistent whether macro-location indices are included or not. A similar trend is observed for all model coefficients when comparing the third and fourth models, with the one notable exception being the coefficient associated with the macro-location indicator. This suggests that part of the index is explained by agglomeration-specific VC. The fully specified model, shown in column (4), indicates that the model explains up to 81% of the variability in sales transactions, and, relevant to this study, indicates that Visual Capital has a positive influence on price in the Lausanne agglomeration used for the two design scenarios.

*Table 1: Regression results across four models, where the dependent variable is the natural logarithm of the transacted price. Column (1) presents the regression results of the model that includes only the variable of interest, visual capital (VC). Column (2) presents the results of the model containing VC conditional on the agglomeration. Column (3) incorporates the fully specified model with an unconditional VC . Column (4) presents results for the fully specified model with VC conditioned on lake-side agglomeration. Robust standard errors are shown in brackets and statistical significance is denoted at the following levels ***p < 0.01, **p < 0.05, *p < 0.1.*

| Parameters | (1) | (2) | (3) | (4) |
|---|---|---|---|---|
| $Intercept$ | -50.62*** [3.85] | -46.92*** [3.48] | -55.47*** [3.03] | -61.87*** [2.79] |
| $VisualCapital(VC)$ | 1.62*** [0.04] | - | 0.27*** [0.03] | - |
| VC: [Biel/Bienne] | - | 1.39*** [0.04] | - | 0.29*** [0.03] |
| VC: [Genève] | - | 1.54*** [0.04] | - | 0.37*** [0.03] |
| VC: [Lausanne] | - | 1.5*** [0.04] | - | 0.35*** [0.03] |
| VC: [Luzern] | - | 1.5*** [0.04] | - | 0.34*** [0.03] |
| VC: [Neuchâtel] | - | 1.42*** [0.04] | - | 0.3*** [0.03] |
| VC: [Thun] | - | 1.42*** [0.04] | - | 0.32*** [0.03] |
| VC: [Vevey–Montreux] | - | 1.49*** [0.04] | - | 0.33*** [0.02] |
| VC: [Zug] | - | 1.59*** [0.04] | - | 0.4*** [0.03] |
| VC: [Zürich] | - | 1.54*** [0.04] | - | 0.39*** [0.03] |
| $Year_{TRANSACTION}$ | 0.03*** [0.0] | 0.03*** [0.0] | 0.03*** [0.0] | 0.03*** [0.0] |
| $\log Volume$ | - | - | 0.37*** [0.02] | 0.39*** [0.02] |
| $N.Rooms$ | - | - | 0.06*** [0.0] | 0.05*** [0.0] |
| $Condition$ | - | - | 0.05*** [0.0] | 0.05*** [0.0] |
| $FitoutStandard$ | - | - | 0.16*** [0.0] | 0.14*** [0.0] |
| $\log Distance_{SEA}$ | - | - | -0.05*** [0.0] | -0.08*** [0.0] |
| $Age$ | - | - | 0.15 [0.33] | 0.34 [0.3] |
| $\log PlotArea$ | - | - | 0.13*** [0.01] | 0.18*** [0.01] |
| $\log MacroLocation$ | - | - | 0.67*** [0.01] | 0.4*** [0.01] |
| Adj. R-squared | 0.21 | 0.36 | 0.76 | 0.81 |
| Observations | 7651 | 7651 | 7651 | 7651 |
| **R-squared** | **0.21** | **0.36** | **0.76** | **0.81** |

Figure 3 provides an illustration of the varying price effect of VC across lakeside agglomerations. Additionally, it shows the range of VC values used to train model 4. Lausanne, displayed in red, has the fourth largest coefficient, and third largest maximum VC range.

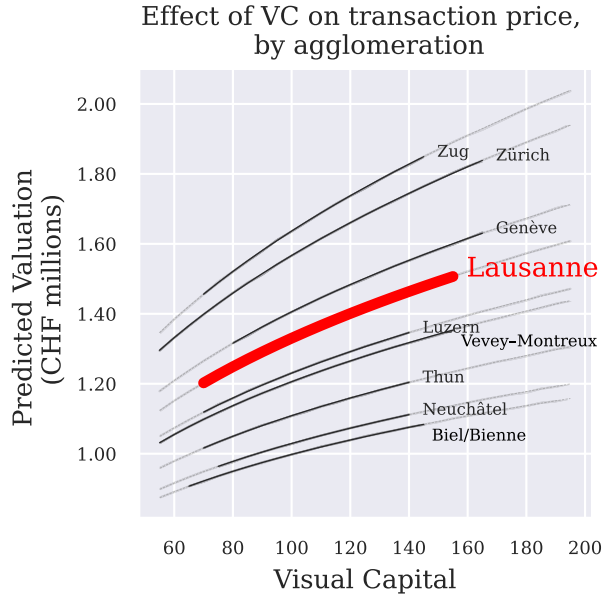

*Figure 3: Price Effect of Visual Capital by agglomeration while holding the other model parameters constant. The black line represents the actual range of values used during model training. For comparison, Lausanne (the agglomeration used in the case studies), is shown in red.*

The remainder of the paper describes results from two case studies which apply the fitted pricing model: (1) the local visual impact on neighboring buildings due to a single proposed development in central Lausanne, and (2) the visual capital at risk due to localized up-zoning in the commune of Saint-Sulpice.

## 4.2 Single Hazard

The proposed 15-story Rasude development negatively impacts the visual metrics of 50% of buildings within a 500m radius. To understand the extent of visual impact, the largest relative loss across all view-metrics – the maximally exposed view metric (MEVM) – is computed and summarized across all buildings. Figure 4 illustrates that approximately 65% of the buildings have a MEVM of less than 1% relative change and that impact on both MEVM and prices are highly concentrated. Figure 4b shows among the most common negatively exposed view-metrics is sky-exposure, proportion of distant views (>1km), as well as the maximum share on water-body, industrial complexes, and nature, with each being impacted in some capacity for 20% of the sampled buildings. As expected, there are positive gains in view-metrics related to façade and near distance obstructions. Note, that the largest relative losses are for scarce view-metrics, such as distant views and water-bodies; and the largest absolute changes are for more abundant view-metrics, such as sky-exposure.

    To understand how price impact is distributed, I compare the aggregate valuations across all buildings in the sample region. Figure 4c illustrates that 44% of aggregate value lost is held by only 4% of the neighbors. They individually have losses greater than 5%, where the most price sensitive building lost 16% of its original valuation. Nearly 40% of the building stock account for the majority or 63% of aggregate value lost, where each individual loss is between 0-5% of the initial valuation (Figure 4c). Interestingly, 8% of the buildings sampled gain value as a consequence of the change in urban form. An analysis of this building subset shows that they benefit from the development of the sky-line. Specifically, the minimal obstruction of positive views with an increase in the visual complexity of the panorama, results in a gain in visual capital, and, in turn predicted price (Figure 4).

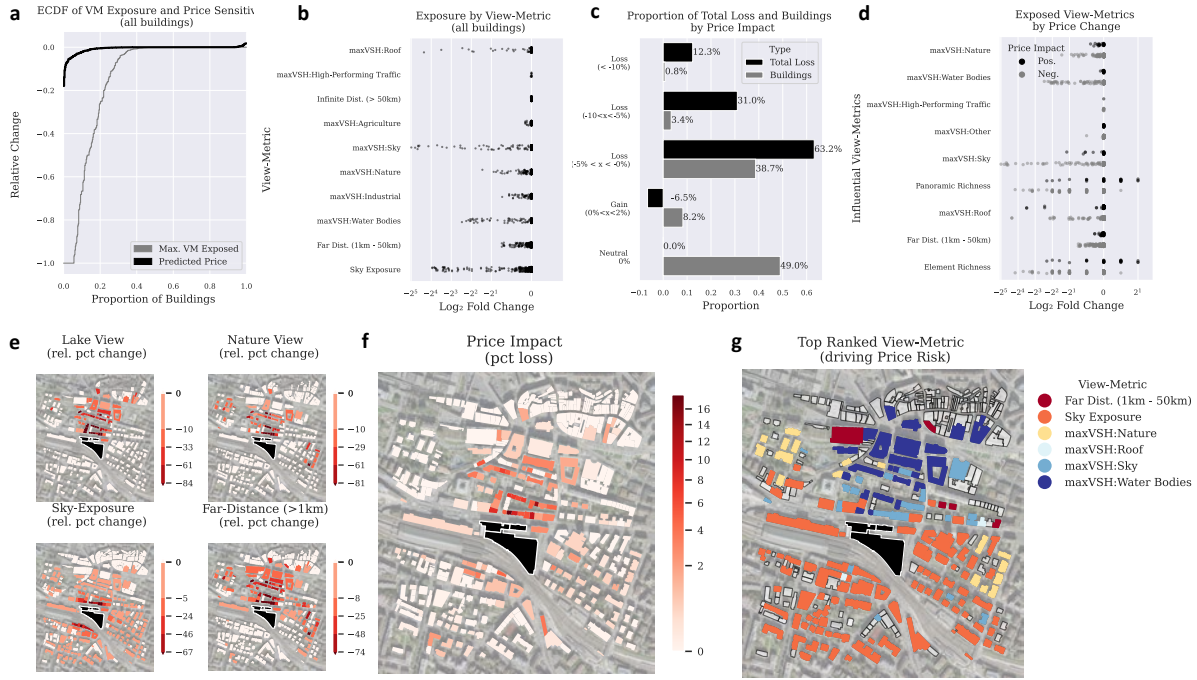

*Figure 4: (a) ECDF of relative change of maximally exposed view-metric and predicted price for all buildings in the sample region. Maximum visual impact is defined as the maximum relative change across a building's vector of view-metrics between the two design scenarios. Heavily skewed distribution indicates concentrated losses. (b) Summary of log fold changes of view-metrics for all buildings between two design scenarios, the design scenario with the proposed development versus the baseline, as-built condition. (c) Barplot of the proportion of aggregate value lost for each level of price sensitivity, relative to the sample share of corresponding buildings. (d) Dot plot comparing the exposed view-metrics of buildings with Positive and Negative price sensitivity, suggesting some building benefits from additions to the 'Sky-Line' (Panoramic and Element Richness). (e) Series of effect-size maps following the developed method: 4-view metrics (lake-view, nature view, far-distance, and sky-exposure). (f) Effect size map of the predicted price impact the proposed development site has on neighboring buildings (development shown in black). (g) Top ranked view-metric contributing to price risk at a given building.*

The spatial distribution of price impact expands radially from the proposed development site, yet, a disproportionate share of the aggregate losses is held by the adjacent neighbors to the north (Figure 4f). Figure 4e shows the spatial distribution of effect size for the sample regions for individual view metrics. As expected, the spatial pattern of exposure varies by view-metrics; contingent on the location and abundance of landcover elements. For instance, impacted lake-views are exclusively to the north of the development site (Lake Geneva is directly south on the development site); and impacted nature views are additionally found in pockets in the east and north west of the sample region (Jura Mountains to the west, Swiss Alps to the east, and French Alps to the South); whereas a radial impact zone appears for sky-exposure. Using the weighting importance of view-metric in estimating visual capital, Figure 4g depicts the metric most responsible for driving the change in predicted price. For examples changes to desirable visual qualities – e.g. lake-views, are the driving determinant for the high price impact region.

### 4.3 Multi Hazard

Results from the regional simulation of up-zoning each building in the commune of Saint-Sulpice by one additional floor confirm that neighboring buildings face devaluation risk caused by nearby developments, with estimates as high as 5% of value lost for individual buildings. Despite the predicted price exposure to local effect (ELE) of individual buildings, the direct effect (DE) of most simulated single-story additions results in aggregate housing price gain even after accounting for the cumulative local effects (CLE).

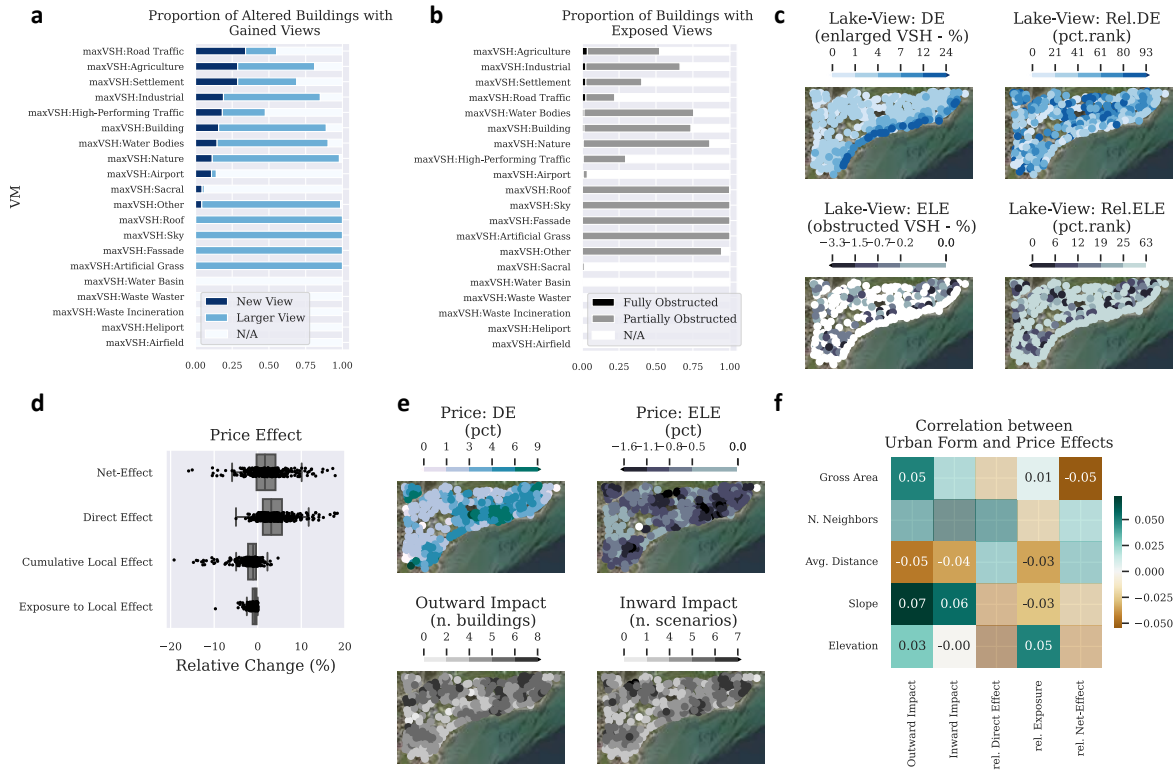

*Figure 5 Barplot of the share of (a) altered buildings with newly gained or enlarged existing views (direct effects) of specific landcover elements. (b) Share of exposed buildings with partial or full obstruction by view metric. (c) Spatial distribution of Lake-View direct effects, relative direct effect ranks, local exposure, and relative exposure ranks. (d) Boxplot of the Price effect across all design scenario; including direct (DE), cumulative local (CLE), exposure (ELE), and net effect. (e) Spatial distribution of Price Effects (DE and ELE); illustrating spatial variability of the number of impacted building to a specific hazard and the count of hazards a building is exposed to (Visual Risk). (f) Correlation plot of price effect metric and Urban and Environmental form attributes, with correlation values shown for significant values $p < .05$.*

For direct effects (DE) at the site of alteration, the majority of absolute gain is defined by the enlargement of already visible abundant landcovers: vegetation, sky, mid-distance (Figure 5a). The most common new views, or the landcover elements not visible prior to the alteration, are local roads, industrial areas, and agriculture. Of the exposed views: landcover area identified as local roads, industrial areas, and agriculture are most at risk of complete obstruction. Figure 5b shows that abundant view-metrics account for the majority of partial obstructions: including sky exposure and vegetation. The average loss of the maximally exposed view-metric (MEVM) is 10%, whereas 15% of the sample risks completely losing its maximally exposed view-metric. Additionally, potential visual impact is a function of the development's location. Despite the large DE and CLE values, the majority of change is explained by abundant and negative sentiment; desirable views account for smaller proportion due to their scarce nature and as such exhibit spatial patterns in change. Figure 5c maps the spatial distribution of lake-view changes, showing that buildings along the shoreline enlarge their lake-view the most, and inland buildings have the greatest relative gain. Whereas the exposed lake-views are distributed in two distinct pockets on the west and eastern edges of the commune. Thus, even though changes to individual view-metrics provide insight to the extent of exposure, they alone do not describe the overall impact, as the importance of the metric, or sensitivity to change in value, have not been accounted for. The following section describe results in terms of price, which can be thought of as weighted combination of building performance metrics according to the learned market preferences.

### 4.3.1 Price Risk

The automated design appraisal model (ADA) captures the price effect with respect to a given design change. The average direct effect of single floor additions in Saint-Sulpice result in a 4.4% price improvement, whereas the highest ranked building gains 12.5%. Interestingly, the rank of direct effect, or price gain, is weakly negatively correlated to both the rank cumulative local effects (CLE), i.e. social cost imposed, and exposure to local effects (ELE), or price vulnerability to local changes. Considering price change, the cumulative local effect (CLE) remains small compared to the direct effect (DE), i.e. price effect at the point of modification effect. Figure 5d shows that the vast majority of design scenarios are a net-positive for Saint-Sulpice, with only 5 locations where the DE is less than the cost imposed through visual obstructions to neighboring buildings. Figure 5f maps the spatial distribution of price changes, showing several distinct pockets of buildings along the shoreline with the largest relative gain in value. Yet, the spatial distribution of value at risk does not follow the same spatial pattern, with multiple clusters forming for both inward and outward impact Figure 5e. To examine the apparent spatial pattern found in the analysis, I subsequently examine the relationship of price impact with characteristics of the urban and natural form. Figure 5f illustrates the correlation between the price effect metrics and urban environmental form metrics, such as slope, building density, elevation, and spread. Although correlations are weak, they are significant, suggesting that, on average, urban form influences the price effect of simulated modifications. For example, buildings in low-density areas, greater distance to neighbors, correlates with larger benefits to alterations (DE) and smaller value at risk (ELE).

It is additionally useful to understand the individual factors, in this case the view-metrics, driving the spatial patterns in both price gain (DE) and value at risk (ELE) within the region. To examine this, Figure 6 depicts the change in a common set of view-metrics from two properties: one from a region of high price gain (DE), and another from a region of high price vulnerability (ELE). The first property (EGID 796374), sees benefits from the alteration, such as a new lake-view, and increased view of natures with little risk to its view-metrics from neighboring local development. However, the second property (EGID), has its 4% visual share of the lake at risk of obstruction due to a single neighbor's alterations, moreover its view share of nature is at risk of obstruction by multiple potential local developments.

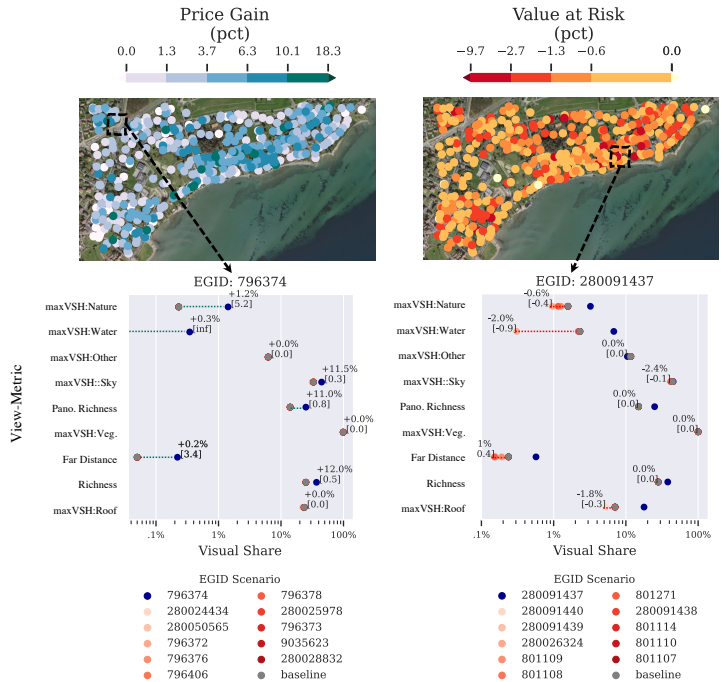

*Figure 6: Change in top weighted view-metrics for 2 separate properties, EGID 796374 from a region of high price gain and EGID 280091437 representing a region from high price risk. Points in grey represent the values for the reference scenario, or as-built condition; Blue points represent visual share value after alteration at the property; and the set of Red points represent the values after the modification it's set of neighbors. Change in visual share (%) of view-metric is expressed for DE with green dashed line, and ELE with red dashed line. The relative change in listed in brackets. For example, the maximum visual share of water (maxVSH:Water) for EGID 280091437 in the reference scenario is ~2.3%; whereas it drops to .3% when EGID 280026324 builds up an addition floor, and rises to 8% when EGID 280091437 itself build up an additional floor.*

Summarizing the metrics driving price gain and risk, Figure 7 depicts the ranked feature importance for both price gain and risk across all buildings. For Saint-Sulpice, maxVSH of water-bodies is the primary determinant of price gain for most site alterations (Figure 7a,b), and is a within the top 3 factors for nearly 60% of the building stock (Figure 7c). For price risk, due to the long coast, proportionately few buildings have exposed lake-views, thus metrics related to sky exposure, such as the maximum visual share of sky, are more commonly the primary determinant of price risk to single-story up-zoning in Saint-Sulpice (Figure 7d,f), with exception of properties along the coast, where the gained façade (within Neutral Sentiment index) and lost view of roofs play a bigger role (Figure 7e).

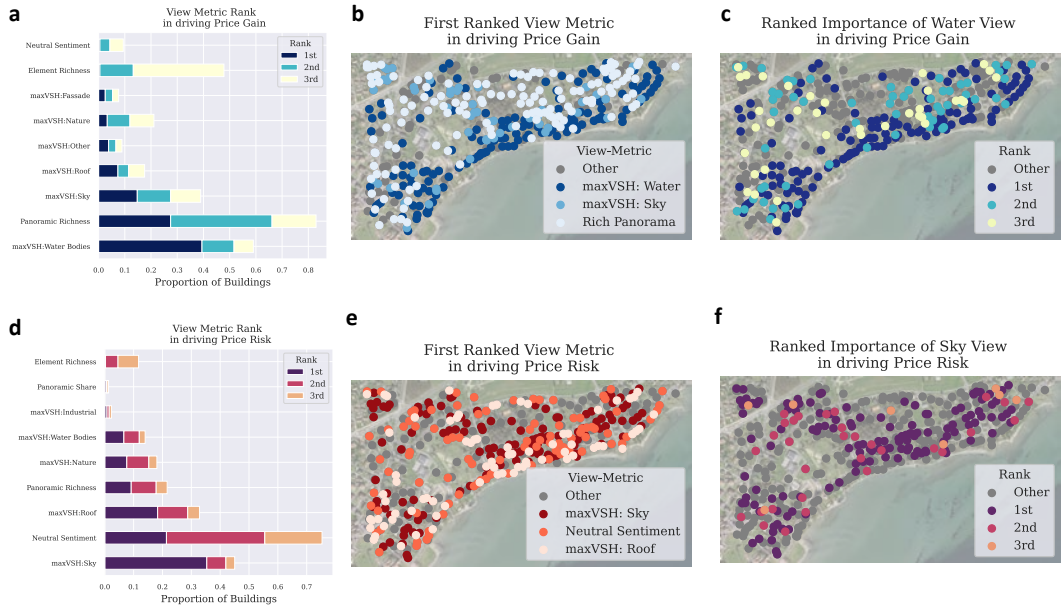

*Figure 7: Predictive importance rank of view-metrics in driving (a) price gain and (d) price risk as a proportion of buildings. (b+e) Maps illustrate the spatial distribution of the first ranked view-metric, and the spatial distribution of (c +f) of the view-metric with largest impact on price: (c) maximum visual share of water and (f) and maximum visual share of sky*

## 5 Discussion

Integrating large scale geometric computing with econometric methods offers an opportunity to infer the price effect of a proposed design alteration. Hence: this paper extends the literature in two ways: First, it presents a novel approach to estimate the financial value of procedurally-generated designs, which I refer to as Automated Design Appraisal (ADA). Second, it incorporates the ADA algorithm within a 3D urban design simulation to measure devaluation risks with respect to design changes. Further, focusing on the urban scale enables the quantification not just of the benefit of a given development at the site of modification, but also of the local vulnerability to the proposed development, i.e. the cost imposed on neighbors by the point of modification.

ADA relies on the assumption that the price of a building is the weighted sum of its individual attributes, also known as the hedonic price theory[29]. Though, unlike the vast hedonic pricing literature, this study takes the additional step to apply the fitted models to newly generated building and urban designs. Utilizing computational design and large-scale 3D geodatabases, the as-built city model is systematically perturbed by altering design parameters, thereby creating new urban scenarios. A subsequent analysis of the price changes relative to the initial as-built conditions help to confirm that urban land-use change has localized impact affecting nearby neighbors to a greater extent. An important aspect of this approach is its interpretability. It is possible to not only explore the impact at the point of modification and nearby buildings; but also understand the determinants of the underlying risk. The latter is achieved by investigating the persistence of specific exposed building attributes– such as lake-view, nature-view, sky-exposure, etc.- and the aggregate valuation at risk due to these specific exposures. A common objection to local development is the immediate impact on visual landscape, which supports the use of Visual Capital in this study; however, future studies may extend this method by focusing on other environmental attributes, including noise pollution, thermal comfort, and air-quality, which are commonly used to raise objection to developments by the NIMBY movement, or more generally communities opposed to local development. Thus, this approach may prove beneficial to local communities interested in quantifying and communicating the visual and environmental impact in terms of local real estate valuation.

A streamlined method to infer the price of computer-generated designs may provide further benefits: Generative Design tools may particularly benefit from ADA. Generative Design in architecture is an iterative design process that outputs feasible building designs under specified optimization functions. The proposed algorithm, ADA enables two new types of objective functions for architectural design optimization. First is optimizing valuation, whereby converting building attributes into monetary terms allows Generative Design procedures to (1) quantify the importance of non-market goods, e.g. environmental quality such as the view, and (2) weigh tradeoffs between seemingly disparate building attributes (e.g. visual quality and programming). The second opportunity, is optimizing for social cost or the cost imposed on its nearby neighbors. It can be reasoned that developments which minimize the localized cost (whether gross cost or count of neighbors negatively impacted) also minimize the risk of local opposition. Evaluating the distribution effects- both direct and localized cost- could be particularly useful for urban planners and property developers who perform pre-development site selection and feasibility studies.

Although the method could be useful for both design optimization and distribution effect exploration, the approach does have its limitations. As with all hedonic analyses, inference is dependent on the specified model. Given that building attributes are highly correlated, results and parameter selection should be approached with care and scrutinized to assure meaningful interpretation. In this paper I study the distribution of price effects of urban development on visual impact. Thus, our results communicate price effects due to visual changes and ignore the changes from other environmental and economic changes that may arise simultaneously. Building upon this work, future studies can incorporate distribution effects stemming from changes to other environmental quality indicators: including noise pollution, thermal comfort, and air-quality.

Lastly, the measures of exposure and sensitivity are ultimately derived from 3D city models. Thus, the level of detail of the underlying 3D model will define the resolution of the building performance metrics. That is, information at a higher fidelity than that of the 3D model will not be included in the evaluation. For instance, building facades in this study are all considered to be the same, ignoring differences in construction materials and textures; lakes are also considered the same, ignoring difference in pollution and geometry. To build upon this limitation, future studies may incorporate images as a way to improve the fidelity of the automated visual impact assessment.

# 6 Conclusion

Economic performance objectives within Architectural Design Optimization have remained challenging to implement, and have thus far been limited to cost minimization, ignoring economic preferences. In this paper, I introduce a novel approach to infer the financial value of a generated design. The proposed Automated Design Appraisal produces building-level price predictions using local-scale environmental performance simulations. Further, I integrate the algorithm within a visual impact framework to understand the property value at risk due to local development.

Results from an impact assessment of a single proposed urban development indicate (1) losses are concentrated to neighbors closest to the point of modification and (2) a subset of buildings benefit from the development of the sky-line, confirming previous findings. Findings from a regional assessment show potential impact -both direct effects and localized costs- are a function of the local urban and environmental form. The spatial pattern of exposure varies by view-metrics; contingent on the location and abundance of landcover elements. Yet, despite the devaluation risk to individual properties, moderate urban development (single-story up-zoning) is estimated to yield aggregate price benefits to low-density regions.

Automated Design Appraisal provides a scalable approach to incorporate economic performance within Architectural Design Optimization procedures. Doing so, enables evaluating generative urban design procedures with respect to both the (1) predicted price and (2) devaluation risk imposed on nearby neighbors. The approach enables future studies to integrate devaluation risk within automated real estate valuation models, reveal mispriced real estate with respect to their local exposure, and aid planners in local zoning and investment decisions.

# References

[1] P. Mayencourt and C. Mueller, "Structural Optimization of Cross-laminated Timber Panels in One-way Bending," *Structures*, vol. 18, pp. 48–59, Apr. 2019, doi: 10.1016/j.istruc.2018.12.009.

[2] J. Natanian and T. Auer, "Beyond nearly zero energy urban design: A holistic microclimatic energy and environmental quality evaluation workflow," *Sustainable Cities and Society*, vol. 56, p. 102094, May 2020, doi: 10.1016/j.scs.2020.102094.

[3] J. Gagne and M. Andersen, "A generative facade design method based on daylighting performance goals," *Journal of Building Performance Simulation*, vol. 5, no. 3, pp. 141–154, May 2012, doi: 10.1080/19401493.2010.549572.

[4] J. Natanian and T. Wortmann, "Simplified evaluation metrics for generative energy-driven urban design: A morphological study of residential blocks in Tel Aviv," *Energy and Buildings*, vol. 240, p. 110916, Jun. 2021, doi: 10.1016/j.enbuild.2021.110916.

[5] Z. Shi, J. A. Fonseca, and A. Schlueter, "A review of simulation-based urban form generation and optimization for energy-driven urban design," *Building and Environment*, vol. 121, pp. 119–129, Aug. 2017, doi: 10.1016/j.buildenv.2017.05.006.

[6] R. E. Weber, C. Mueller, and C. Reinhart, "Automated floorplan generation in architectural design: A review of methods and applications," *Automation in Construction*, vol. 140, p. 104385, Aug. 2022, doi: 10.1016/j.autcon.2022.104385.

[7] R. F. M. Ameen, M. Mourshed, and H. Li, "A critical review of environmental assessment tools for sustainable urban design," *Environmental Impact Assessment Review*, vol. 55, pp. 110–125, Nov. 2015, doi: 10.1016/j.eiar.2015.07.006.

[8] D. Elshani, "Measuring Sustainability and Urban Data Operationalization - An integrated computational framework to evaluate and interpret the performance of the urban form.," in *A. Globa, J. van Ameijde, A. Fingrut, N. Kim, T.T.S. Lo (eds.), PROJECTIONS - Proceedings of the 26th CAADRIA Conference - Volume 2, The Chinese University of Hong Kong and Online, Hong Kong, 29 March - 1 April 2021, pp. 407-416*, CUMINCAD, 2021. Accessed: Sep. 23, 2023. [Online]. Available: https://papers.cumincad.org/cgi-bin/works/paper/caadria2021_391

[9] S. Law, B. Paige, and C. Russell, "Take a Look Around: Using Street View and Satellite Images to Estimate House Prices," *ACM Trans. Intell. Syst. Technol.*, vol. 10, no. 5, pp. 1–19, Nov. 2019, doi: 10.1145/3342240.

[10] W. Y. Chen, X. Li, and J. Hua, "Environmental amenities of urban rivers and residential property values: A global meta-analysis," *Science of The Total Environment*, vol. 693, p. 133628, Nov. 2019, doi: 10.1016/j.scitotenv.2019.133628.

[11] X. Dai, D. Felsenstein, and A. Y. Grinberger, "Viewshed effects and house prices: Identifying the visibility value of the natural landscape," *Landscape and Urban Planning*, vol. 238, p. 104818, Oct. 2023, doi: 10.1016/j.landurbplan.2023.104818.

[12] H. H. Rong, J. Yang, M. Kang, and A. Chegut, "The Value of Design in Real Estate Asset Pricing," *Buildings*, vol. 10, no. 10, Art. no. 10, Oct. 2020, doi: 10.3390/buildings10100178.

[13] J. Yang, H. Rong, Y. Kang, F. Zhang, and A. Chegut, "The financial impact of street-level greenery on New York commercial buildings," *Landscape and Urban Planning*, vol. 214, p. 104162, Oct. 2021, doi: 10.1016/j.landurbplan.2021.104162.

[14] I. Turan, A. Chegut, D. Fink, and C. Reinhart, "The value of daylight in office spaces," *Building and Environment*, vol. 168, p. 106503, Jan. 2020, doi: 10.1016/j.buildenv.2019.106503.

[15] I. Turan, A. Chegut, D. Fink, and C. Reinhart, "Development of view potential metrics and the financial impact of views on office rents," *Landscape and Urban Planning*, vol. 215, p. 104193, Nov. 2021, doi: 10.1016/j.landurbplan.2021.104193.

[16] A. Baranzini and C. Schaerer, "A sight for sore eyes: Assessing the value of view and land use in the housing market," *Journal of Housing Economics*, vol. 20, no. 3, pp. 191–199, Sep. 2011, doi: 10.1016/j.jhe.2011.06.001.

[17] F. Jiang *et al.*, "Generative urban design: A systematic review on problem formulation, design generation, and decision-making," *Progress in Planning*, p. 100795, Jul. 2023, doi: 10.1016/j.progress.2023.100795.

[18] W. H. Ko *et al.*, "Window View Quality: Why It Matters and What We Should Do," *LEUKOS*, vol. 18, no. 3, pp. 259–267, Jul. 2022, doi: 10.1080/15502724.2022.2055428.

[19] M. Roth, S. Hildebrandt, U. Walz, and W. Wende, "Large-Area Empirically Based Visual Landscape Quality Assessment for Spatial Planning—A Validation Approach by Method Triangulation," *Sustainability*, vol. 13, no. 4, Art. no. 4, Jan. 2021, doi: 10.3390/su13041891.

[20] W. A. Fischel, "Why Are There NIMBYs?," *Land Economics*, vol. 77, no. 1, pp. 144–152, Feb. 2001, doi: 10.2307/3146986.

[21] A. R. Swietek and M. Zumwald, "Visual Capital: Evaluating building-level visual landscape quality at scale," *Landscape and Urban Planning*, vol. 240, p. 104880, Dec. 2023, doi: 10.1016/j.landurbplan.2023.104880.

[22] R. Baušys and I. Pankrašovaite, "Optimization of architectural layout by the improved genetic algorithm," *Journal of Civil Engineering and Management*, vol. 11, no. 1, Art. no. 1, Mar. 2005, doi: 10.3846/13923730.2005.9636328.

[23] D. Nagy, L. Villaggi, and D. Benjamin, "Generative urban design: integrating financial and energy goals for automated neighborhood layout," in *Proceedings of the Symposium on Simulation for Architecture and Urban Design*, in SIMAUD '18. San Diego, CA, USA: Society for Computer Simulation International, Jun. 2018, pp. 1–8.

[24] T. G. Thibodeau, "Estimating the Effect of High-Rise Office Buildings on Residential Property Values," *Land Economics*, vol. 66, no. 4, pp. 402–408, 1990, doi: 10.2307/3146622.

[25] N. C. Brown, "Design performance and designer preference in an interactive, data-driven conceptual building design scenario," *Design Studies*, vol. 68, pp. 1–33, May 2020, doi: 10.1016/j.destud.2020.01.001.

[26] T.-K. Wang and W. Duan, "Generative design of floor plans of multi-unit residential buildings based on consumer satisfaction and energy performance," *Developments in the Built Environment*, vol. 16, p. 100238, Dec. 2023, doi: 10.1016/j.dibe.2023.100238.

[27] L. Villaggi, J. Stoddart, D. Nagy, and D. Benjamin, "Survey-Based Simulation of User Satisfaction for Generative Design in Architecture," in *Humanizing Digital Reality: Design Modelling Symposium Paris 2017*, K. De Rycke, C. Gengnagel, O. Baverel, J. Burry, C. Mueller, M. M. Nguyen, P. Rahm, and M. R. Thomsen, Eds., Singapore: Springer, 2018, pp. 417–430. doi: 10.1007/978-981-10-6611-5_36.

[28] S. Fifer, J. Rose, and S. Greaves, "Hypothetical bias in Stated Choice Experiments: Is it a problem? And if so, how do we deal with it?," *Transportation Research Part A: Policy and Practice*, vol. 61, pp. 164–177, Mar. 2014, doi: 10.1016/j.tra.2013.12.010.

[29] S. Rosen, "Hedonic Prices and Implicit Markets: Product Differentiation in Pure Competition," *Journal of Political Economy*, vol. 82, no. 1, pp. 34–55, 1974.

[30] J. Stroebel and J. Wurgler, "What do you think about climate finance?," *Journal of Financial Economics*, vol. 142, no. 2, pp. 487–498, Nov. 2021, doi: 10.1016/j.jfineco.2021.08.004.

[31] R. Holtermans, D. Niu, and S. Zheng, "Quantifying the Impacts of Climate Shocks in Commercial Real Estate Market." Rochester, NY, Oct. 10, 2022. doi: 10.2139/ssrn.4276452.

[32] F. Ortega and S. Taspınar, "Rising sea levels and sinking property values: Hurricane Sandy and New York's housing market," *Journal of Urban Economics*, vol. 106, pp. 81–100, Jul. 2018, doi: 10.1016/j.jue.2018.06.005.

[33] A. Ouazad and M. Kahn, "Mortgage Finance in the Face of Rising Climate Risk," National Bureau of Economic Research, Cambridge, MA, w26322, Sep. 2019. doi: 10.3386/w26322.

[34] L. Barrage and J. Furst, "Housing investment, sea level rise, and climate change beliefs," *Economics Letters*, vol. 177, pp. 105–108, Apr. 2019, doi: 10.1016/j.econlet.2019.01.023.

[35] L. A. Bakkensen and L. Barrage, "Going Underwater? Flood Risk Belief Heterogeneity and Coastal Home Price Dynamics," *The Review of Financial Studies*, vol. 35, no. 8, pp. 3666–3709, Aug. 2022, doi: 10.1093/rfs/hhab122.

[36] P. Issler, R. Stanton, C. Vergara-Alert, and N. Wallace, "Mortgage Markets with Climate-Change Risk: Evidence from Wildfires in California." Rochester, NY, Jul. 01, 2020. doi: 10.2139/ssrn.3511843.

[37] P. D. Bates *et al*., "Combined Modeling of US Fluvial, Pluvial, and Coastal Flood Hazard Under Current and Future Climates," *Water Resources Research*, vol. 57, no. 2, p. e2020WR028673, 2021, doi: 10.1029/2020WR028673.

[38] J. D. Gourevitch *et al*., "Unpriced climate risk and the potential consequences of overvaluation in US housing markets," *Nat. Clim. Chang.*, vol. 13, no. 3, Art. no. 3, Mar. 2023, doi: 10.1038/s41558-023-01594-8.

[39] E. Nault, G. Peronato, E. Rey, and M. Andersen, "Review and critical analysis of early-design phase evaluation metrics for the solar potential of neighborhood designs," *Building and Environment*, vol. 92, pp. 679–691, Oct. 2015, doi: 10.1016/j.buildenv.2015.05.012.

[40] P. Florio, G. Peronato, A. T. D. Perera, A. Di Blasi, K. H. Poon, and J. H. Kämpf, "Designing and assessing solar energy neighborhoods from visual impact," *Sustainable Cities and Society*, vol. 71, p. 102959, Aug. 2021, doi: 10.1016/j.scs.2021.102959.

[41] A. Baranzini, J. V. Ramirez, C. Schaerer, and P. Thalmann, "Introduction to this Volume: Applying Hedonics in the Swiss Housing Markets," *Swiss J Economics Statistics*, vol. 144, no. 4, pp. 543–559, Oct. 2008, doi: 10.1007/BF03399265.

[42] A. Baranzini, C. Schaerer, J. V. Ramirez, and P. Thalmann, "Feel it or Measure it - Perceived vs. Measured Noise in Hedonic Models." Rochester, NY, Oct. 01, 2006. Accessed: Jul. 20, 2022. [Online]. Available: https://papers.ssrn.com/abstract=937259

[43] N. C. Inglis, J. Vukomanovic, J. Costanza, and K. K. Singh, "From viewsheds to viewscapes: Trends in landscape visibility and visual quality research," *Landscape and Urban Planning*, vol. 224, p. 104424, Aug. 2022, doi: 10.1016/j.landurbplan.2022.104424.

[44] M. Li, F. Xue, Y. Wu, and A. G. O. Yeh, "A room with a view: Automatic assessment of window views for high-rise high-density areas using City Information Models and deep transfer learning," *Landscape and Urban Planning*, vol. 226, p. 104505, Oct. 2022, doi: 10.1016/j.landurbplan.2022.104505.

[45] D. Cilliers, M. Cloete, A. Bond, F. Retief, R. Alberts, and C. Roos, "A critical evaluation of visibility analysis approaches for visual impact assessment (VIA) in the context of environmental impact assessment (EIA)," *Environmental Impact Assessment Review*, vol. 98, p. 106962, Jan. 2023, doi: 10.1016/j.eiar.2022.106962.

[46] J. Vukomanovic and B. J. Orr, "Landscape Aesthetics and the Scenic Drivers of Amenity Migration in the New West: Naturalness, Visual Scale, and Complexity," *Land*, vol. 3, no. 2, Art. no. 2, Jun. 2014, doi: 10.3390/land3020390.

[47] C. Gaigné, H. R. A. Koster, F. Moizeau, and J.-F. Thisse, "Who lives where in the city? Amenities, commuting and income sorting," *Journal of Urban Economics*, vol. 128, p. 103394, Mar. 2022, doi: 10.1016/j.jue.2021.103394.

[48] D. Djurdjevic, C. Eugster, and R. Haase, "Estimation of Hedonic Models Using a Multilevel Approach: An Application for the Swiss Rental Market," *Swiss J Economics Statistics*, vol. 144, no. 4, Art. no. 4, Oct. 2008, doi: 10.1007/BF03399271.

[49] Federal Office of Topography swisstopo, "swissALTI3D," Federal Office of Topography swisstopo. Accessed: Dec. 05, 2022. [Online]. Available: https://www.swisstopo.admin.ch/en/geodata/height/alti3d.html

[50] Federal Office of Topography swisstopo, "swissBUILDINGS3D 2.0," Federal Office of Topography swisstopo. Accessed: Jun. 09, 2023. [Online]. Available: https://www.swisstopo.admin.ch/en/geodata/landscape/buildings3d2.html

[51] "Vegetation Height Model NFI - 2019 Vegetation Height Model NFI (current) - EnviDat." Accessed: Sep. 10, 2023. [Online]. Available: https://www.envidat.ch/dataset/vegetation-height-model-nfi/resource/d4f64aef-f65e-4070-a661-dac8c49abc69

[52] Intergovernmental Panel On Climate Change (Ipcc), *Climate Change 2022 – Impacts, Adaptation and Vulnerability: Working Group II Contribution to the Sixth Assessment Report of the Intergovernmental Panel on Climate Change*, 1st ed. Cambridge University Press, 2023. doi: 10.1017/9781009325844.

[53] "La Rasude reprend vie au cœur de Lausanne.," La Rasude. Accessed: Oct. 11, 2023. [Online]. Available: https://la-rasude.ch/

[54] "Quartier Rasude à Lausanne – Quinze étages, «c'est un cadeau aux promoteurs»," 24 heures. Accessed: Sep. 11, 2023. [Online]. Available: https://www.24heures.ch/quinze-etages-cest-un-cadeau-aux-promoteurs-147843395226

[55] "L'association Perirasude," Association Perirasude. Accessed: Nov. 09, 2023. [Online]. Available: https://perirasude.com/